\documentclass[10pt,twocolumn,twoside]{IEEEtran}

\usepackage{booktabs}

\usepackage{subfigure}
\usepackage{amssymb}
\usepackage{amsfonts}
\usepackage{amsmath}
\usepackage{cite}
\usepackage[dvips]{graphicx}
\usepackage{color}
\usepackage{comment}
\usepackage{makecell}
\usepackage{multirow}
\usepackage{textcomp,mathcomp}

\begin{document}
\newtheorem{definition}{\it Definition}
\newtheorem{theorem}{\bf Theorem}
\newtheorem{lemma}{\it Lemma}
\newtheorem{corollary}{\it Corollary}
\newtheorem{remark}{\it Remark}
\newtheorem{example}{\it Example}
\newtheorem{case}{\bf Case Study}
\newtheorem{assumption}{\it Assumption}
\newtheorem{property}{\it Property}
\newtheorem{proposition}{\it Proposition}

\newcommand{\hP}[1]{{\boldsymbol h}_{{#1}{\bullet}}}
\newcommand{\hS}[1]{{\boldsymbol h}_{{\bullet}{#1}}}

\newcommand{\ba}{\boldsymbol{a}}
\newcommand{\baq}{\overline{q}}
\newcommand{\bA}{\boldsymbol{A}}
\newcommand{\bb}{\boldsymbol{b}}
\newcommand{\bB}{\boldsymbol{B}}
\newcommand{\bc}{\boldsymbol{c}}
\newcommand{\bcO}{\boldsymbol{\cal O}}
\newcommand{\bh}{\boldsymbol{h}}
\newcommand{\bH}{\boldsymbol{H}}
\newcommand{\bl}{\boldsymbol{l}}
\newcommand{\bm}{\boldsymbol{m}}
\newcommand{\bn}{\boldsymbol{n}}
\newcommand{\bo}{\boldsymbol{o}}
\newcommand{\bO}{\boldsymbol{O}}
\newcommand{\bp}{\boldsymbol{p}}
\newcommand{\bq}{\boldsymbol{q}}
\newcommand{\bR}{\boldsymbol{R}}
\newcommand{\bs}{\boldsymbol{s}}
\newcommand{\bS}{\boldsymbol{S}}
\newcommand{\bT}{\boldsymbol{T}}
\newcommand{\bw}{\boldsymbol{w}}

\newcommand{\balpha}{\boldsymbol{\alpha}}
\newcommand{\bbeta}{\boldsymbol{\beta}}
\newcommand{\bOmega}{\boldsymbol{\Omega}}
\newcommand{\bTheta}{\boldsymbol{\Theta}}
\newcommand{\bphi}{\boldsymbol{\phi}}
\newcommand{\btheta}{\boldsymbol{\theta}}
\newcommand{\bvarpi}{\boldsymbol{\varpi}}
\newcommand{\bpi}{\boldsymbol{\pi}}
\newcommand{\bpsi}{\boldsymbol{\psi}}
\newcommand{\bxi}{\boldsymbol{\xi}}
\newcommand{\bx}{\boldsymbol{x}}
\newcommand{\by}{\boldsymbol{y}}

\newcommand{\cA}{{\cal A}}
\newcommand{\bcA}{\boldsymbol{\cal A}}
\newcommand{\cB}{{\cal B}}
\newcommand{\cE}{{\cal E}}
\newcommand{\cG}{{\cal G}}
\newcommand{\cH}{{\cal H}}
\newcommand{\bcH}{\boldsymbol {\cal H}}
\newcommand{\cK}{{\cal K}}
\newcommand{\cO}{{\cal O}}
\newcommand{\cR}{{\cal R}}
\newcommand{\cS}{{\cal S}}
\newcommand{\dcS}{\ddot{{\cal S}}}
\newcommand{\ds}{\ddot{{s}}}
\newcommand{\cT}{{\cal T}}
\newcommand{\cU}{{\cal U}}
\newcommand{\wt}[1]{\widetilde{#1}}

\newcommand{\mA}{\mathbb{A}}
\newcommand{\mE}{\mathbb{E}}
\newcommand{\mG}{\mathbb{G}}
\newcommand{\mR}{\mathbb{R}}
\newcommand{\mS}{\mathbb{S}}
\newcommand{\mU}{\mathbb{U}}
\newcommand{\mV}{\mathbb{V}}
\newcommand{\mW}{\mathbb{W}}

\newcommand{\uq}{\underline{q}}
\newcommand{\ubq}{\underline{\boldsymbol q}}

\newcommand{\red}[1]{\textcolor[rgb]{1,0,0}{#1}}
\newcommand{\gre}[1]{\textcolor[rgb]{0,1,0}{#1}}
\newcommand{\blu}[1]{\textcolor[rgb]{0,0,1}{#1}}

\title{Towards Agentic AI Networking in 6G: A Generative Foundation Model-as-Agent Approach}

\author{Yong~Xiao, Guangming~Shi, and Ping Zhang 

\thanks{*This work is accepted at IEEE Communications Magazine. Copyright may be transferred without notice, after which this version may no longer be accessible.

Y. Xiao is with the School of Electronic Information and Communications, the Huazhong University of Science and Technology, Wuhan, China 430074, also with the Peng Cheng Laboratory, Shenzhen, China, and also with Pazhou Laboratory (Huangpu), Guangzhou, China
(e-mail: yongxiao@hust.edu.cn).


G. Shi is with the Peng Cheng Laboratory, Shenzhen, China 518055, also with the School of Artificial Intelligence, Xidian University, Xi'an, Shaanxi 710071, China, and also with Pazhou Laboratory (Huangpu), Guangzhou, China
(e-mail: gmshi@xidian.edu.cn).

P. Zhang is with the State Key Laboratory of Networking and Switching Technology, Beijing University of Posts and Telecommunications, Beijing, China 100876 (email: pzhang@bupt.edu.cn).
}
}








\maketitle

\begin{abstract}
The promising potential of AI and network convergence in improving networking performance and enabling new service capabilities has recently attracted significant interest. Existing network AI solutions, while powerful, are mainly built based on the close-loop and passive learning framework, resulting in major limitations in autonomous solution finding and dynamic environmental adaptation. Agentic AI has recently been introduced as a promising solution to address the above limitations and pave the way for true, generally intelligent, and beneficial AI systems. The key idea is to create a networking ecosystem to support a diverse range of autonomous and embodied AI agents in fulfilling their goals. In this paper, we focus on the novel challenges and requirements of agentic AI networking. We propose {\em AgentNet}, a novel framework for supporting interaction, collaborative learning, and knowledge transfer among AI agents. We introduce a general architectural framework of AgentNet and then propose a generative foundation model (GFM)-based implementation in which multiple GFM-as-agents have been created as an interactive knowledge-base to bootstrap the development of embodied AI agents according to different task requirements and environmental features. We consider two application scenarios, digital-twin-based industrial automation and metaverse-based infotainment system, to describe how to apply AgentNet for supporting efficient task-driven collaboration and interaction among AI agents. 

\end{abstract}

\section{Introduction}
\label{Section_Introduction}

The seamless integration of artificial intelligence (AI) and communication networks is widely recognized as a pivotal trend shaping the future of networking systems, including 6G and beyond. 
Future networking systems will be dominated by a large volume of densely deployed AI models and applications in both the physical and virtual worlds, tailored to specific task objectives and requirements across various industries\cite{Yang2022NetMagazine,XY2020Selflearning}. Communication between AI models focuses primarily on collaborative model construction and coordination for specific task objectives, which exhibit fundamentally different requirements compared to existing data-focused communication networks. Major standardization development organizations (SDOs), including 3GPP and ITU-T, are actively working on solutions to improve existing communication networks to accommodate these new requirements. For example, AI capabilities and functions have been incorporated into the Network Data Analytics Function (NWDAF) in the core network architecture since the first version of 5G, i.e., 3GPP Release 15. The ITU-T Focus Group on AI-Native Networks was also established in July 2024 with the main scope of exploring and identifying the fundamental changes required for the future network architecture to harness the full potential of AI. 
The applications of AI technology in the radio access network (RAN) have also attracted significant interest by academia and industry. For example, the AI-RAN Alliance was recently established to accelerate the integration of AI into RAN technologies to improve spectrum utilization and energy efficiency\cite{Habibi2024AIMLinORAN}.

Despite this progress, the current communication networking architecture was designed primarily to passively transport data packets. 
The inherent separation of data transportation and processing functions is still recognized as a significant impediment to supporting efficient, responsive, and secure AI networking. More specifically, the increasing popularity of AI services and applications poses the following novel challenges to existing communication networks:

\noindent{\bf (1) Data traffic generation speed significantly exceeds network capacity:} The network infrastructure deployment speed has long been known to lag behind the growing speed of mobile data traffic. With the increasing popularity of AI-based applications and services, the gap between the volume of worldwide data traffics and the capacity of global network infrastructure will only grow larger.

\noindent{\bf (2) Limited flexibility and adaptability of network AI models cannot meet the increasingly personalized and diversified needs: } Most existing networking AI solutions are built based on closed-loop and passive learning frameworks in which models are trained by historical datasets based on the assumption that the statistical features or patterns of data in the past will remain in future deployments. These solutions are generally difficult to meet the increasingly personalized and diversified needs of users.

\noindent{\bf (3) Growing concerns on user data and model security: } 
  With the increasing integration of AI models and services, including e-personal assistance and e-healthcare, into human life, more and more sensitive data like medical records, financial transactions, or personal communications have been generated by smart devices and UEs. These data can be vulnerable when exposed to communication networks and edge or cloud service providers.

Recently, agentic AI, a novel AI paradigm focusing on developing capabilities of taking autonomous and independent actions and interact with the environment, has been introduced as a promising solution to address the above challenges and 
pay the way for the more flexible, generally intelligent, and secure AI services and applications in the future\cite{Morris2024PositionAGI, Murugesan2025AgenticAI}. 
Most of the existing work focuses on implementing specific agentic AI models and learning frameworks for various application tasks and use cases, while ignoring the fact that the agentic AI system is in essence a communication networking ecosystem consisting of a large number of AI agents interacting, collaborating, and coordinating the necessary information for various goals.

Motivated by the above observation, in this paper, we investigate the agentic AI networking from the perspective of communication networks. We introduce a novel AI-native networking architecture, referred to as AgentNet, for supporting and coordinating the interactions and collaborative efforts of diverse types of agents to achieve their goals effectively.
AgentNet has the potential to fundamentally address the aforementioned challenges. In particular, by allowing the locally collected raw data to be directly applied to train agents, instead of being transported to the network, the network traffic can be significantly reduced. In addition, instead of constructing a single (large) model to achieve a limited set of tasks, AgentNet focuses on implementing a diverse set of agents, each is designed with a unique skillset, to interact and collaborate to meet the personalized and diversified needs of users. Finally, instead of updating models by exposing the raw data or model parameters to the network, agents in AgentNet can learn and update their models through collaboration and interaction, so the data and model security can be preserved. 
In this paper, we introduce the basic concepts, architectures, KPIs, and implementations of AgentNet. Our main contributions are summarized as follows:
\begin{itemize}
    \item[(1)] We introduce AgentNet, a novel agentic AI networking architecture. We describe the unique features, types, and design principles of agents to be supported by AgentNet. The architectural framework of AgentNet, including the key components and main KPIs, is introduced.
    \item[(2)] We propose a generative foundation model (GFM)-based implementation of AgentNet in which multiple GFM-as-agents have been created as an interactive knowledge-base to bootstrap the development of embodied AI agents according to different task requirements and environmental features.
    \item[(3)] 
    We develop an AgentNet prototype and evaluate the performance of AgentNet in two application scenarios, digital-twin-based industrial automation and metaverse-based infotainment systems. 
\end{itemize}

\section{Agentic AI Networking}
As the building blocks of agentic AI networking systems, AI agents are autonomous or semiautonomous intelligent beings supported by necessary knowledge and resources designed to accomplish certain goals in the targeted environment \cite{Murugesan2025AgenticAI}. 
They can be either logical or physical entities, each involves a single AI model or a set of collaborative AI models to perceive, make decisions, and interact with virtual or physical surroundings, and respond accordingly. 
Compared to the traditional AI model-based functional modules, AI agents have the following unique features:

  
  \noindent{\bf (1) Proactive interaction with environments: } Compared to the existing AI models that are passively waiting for input, AI agents can actively seek information, explore possibilities, and initiate actions to interact with their environments that involve other agents and dynamic elements.
  
  \noindent{\bf (2) Goal-oriented self-learning and self-correction: } Compared to the existing AI models that are passively learning from the given training datasets, AI agents can leverage the real-world knowledge and prior experience to self-learn, correct, and refine their decision-making processes and provide fair and balanced responses to achieve different goals with minimized influence from the bias hidden behind the training data. 
  
  \noindent{\bf (3) Life-long learning: } Compared to the existing AI models that are closed-loop and remain fixed after being trained,  AI agents can continuously adapt and improve their performance over time by acquiring new knowledge and skills throughout their operational lifespan.

There are many novel approaches to implementing agents according to different goals and objectives. 
In the rest of this paper, we mainly focus on the following three types of AI agents:


\noindent{\bf (1) Foundation model-as-agent (F-agent): } This corresponds to the agent that is based on models designed for general-purpose and trained on extensive and diverse data sources and is adaptable for a variety of downstream tasks. More specifically, F-agents are often developed based on large foundation models such as the large language model (LLM), the large vision model (LVM), and multimodal models with the ability not only to store and access a broad range of knowledge information, but also to represent various types of skillsets such as reasoning and inferring implicit rules and rationality, and even creating novel contents based on the inferred rules. 

\noindent{\bf (2) Embodied model-as-agent (E-agent): } This corresponds to the agent that is built on task-specific models and can directly interact with specific types of environments. Compared to F-agents, E-agents often rely on light-weight models to ensure responsive interaction and easy deployment in resource-constrained UEs and network equipment. 


\noindent{\bf (3) Hybrid model-as-agents (H-agent): } This corresponds to the agent that requires collaboration between foundation and embodied models. In particular, embodied models can improve the environmental adaptability and applicability of foundation models in many real-world scenarios. Foundation models can also provide the knowledge-base and skillset for embodied models when being deployed in unknown environments. H-agents with integrated foundational and embodied models offer powerful approaches in creating more adaptable, flexible, and generally intelligent systems that can interact effectively with and understand complex environments. 


In this paper, we focus on the communication requirements and architecture for supporting the networking of AI agents.



More specifically, we define {\em AgentNet} as a specialized networking system designed to facilitate efficient information exchange, action coordination, and knowledge transfer to achieve specific goals among heterogeneous agents including F-, E-, and H-agents. Compared to the existing data-oriented networks, AgentNet poses the following three paradigm shifts in design goal, network management, and performance optimization and evaluation solutions:

  
  \noindent{\bf (1) From data-focused to goal-oriented:} Compared to the traditional data-oriented communication networks focusing primarily on transmission and accurate delivery of data packets, AgentNet prioritize the needs of collaborative model construction and joint decision making for achieving specific goals. This results in fundamentally different requirements on the information process, transportation, and evaluation of delivery results\cite{XY2023iSAN}. 


  
  \noindent{\bf (2) From central control to decentralized autonomy: } Since AgentNet interconnects a large number of autonomous agents, the existing centralized management-based networking architecture, reliant on uploading local sensing data to edge or cloud servers for processing, suffer from inefficiencies, reliability issues, and security vulnerabilities. AgentNet should therefore be decentralized in nature, prioritizing edge coordination and knowledge transfer among relevant agents. 

  
  \noindent{\bf (3) From single-resource-focused to multi-resource-related performance optimization: } While the performance of the existing data-focused communication systems primarily relies on spectrum efficiency, the performance of AgentNet is directly or indirectly linked to a diverse set of resources, including accuracy of the local sensors, computational power, storage capacity, and algorithmic design. This means that the performance evaluation metrics of AgentNet should also include non-data-oriented metrics such as environment-adaptability, autonomous levels, etc., as will be discussed in the next section.

As the number of agents continues to increase exponentially, there is a critical need to establish novel architectural frameworks to ensure the sustainable development and scalability of AgentNet. 

\begin{figure}
\centering
\includegraphics[width=3.5 in]{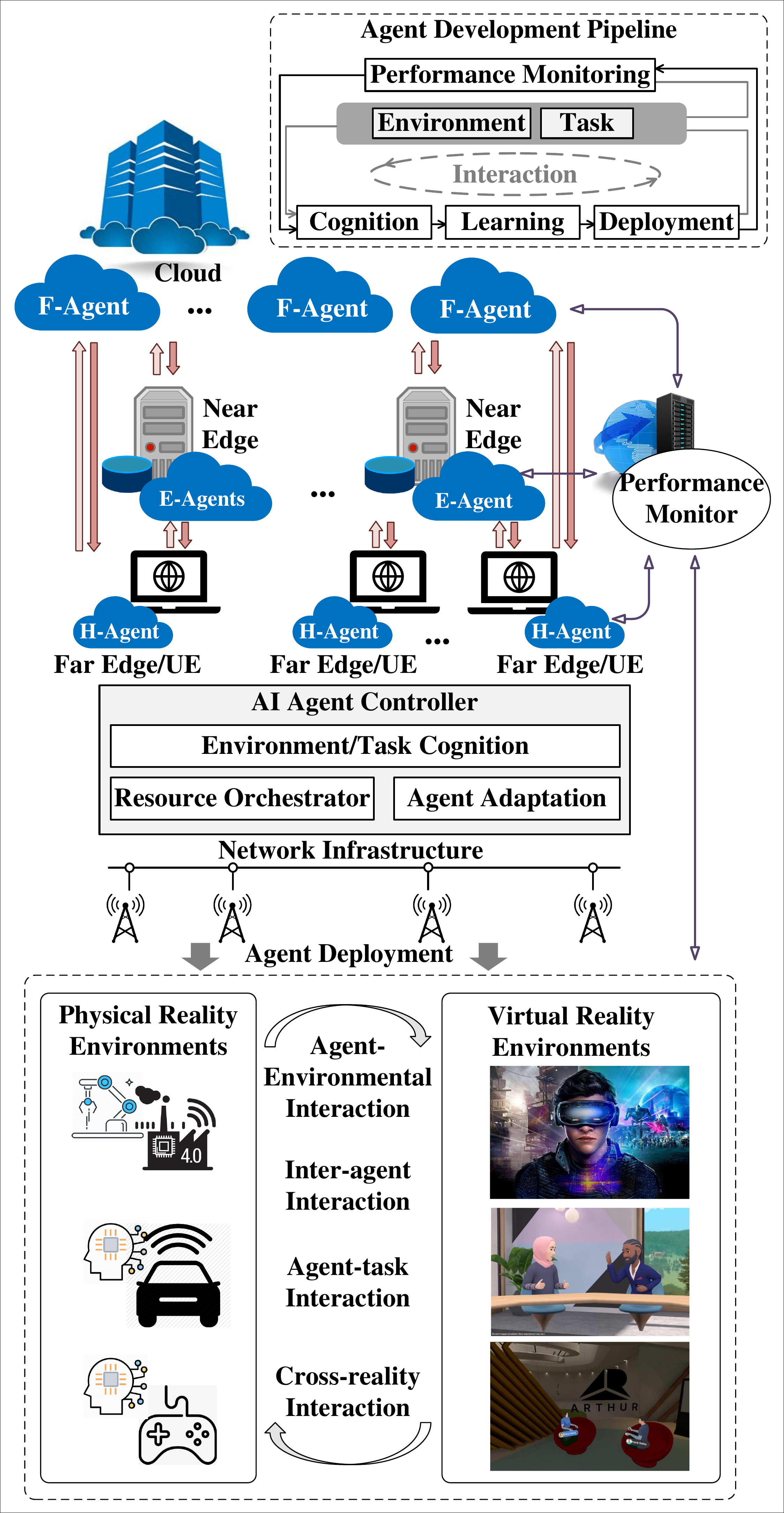}
\caption{{The agent development pipeline (top-right subplot) and the general architectural framework of AgentNet (bottom subplot) including {\em infrastructure} (cloud data center, edge servers, UEs, and network infrastructure), {\em environment} (physical reality and virtual reality environments), {\em application tasks and goals}, {\em agents} and {\em agent controller}.}}
%
\label{Figure_GeneralAgentNetArchitecture}
\end{figure}

\section{Architectural Framework and KPIs}

\subsection{Architectural Framework}

A general AgentNet architecture, as presented in Fig. \ref{Figure_GeneralAgentNetArchitecture}, consists of the following key components:


\noindent{\bf (1) Infrastructure:} includes the hardware infrastructure, such as cloud and edge computing and storage systems and communication networks that interconnect different agents and enable their access to the resources stored at the cloud and edge, as well as the so-called {\em agent infrastructure} including the software platforms of agents and sources of high-quality expert datasets, accumulated skillsets, and knowledge of the world models that are accessible for the agents.


\noindent{\bf (2) Environment:} includes both physical and virtual environments where agents can interact. The agent can develop various skillsets and learn how to behave to achieve their goals by adopting the standard trail-and-correct approach under different environments. Generally speaking, interaction in physical environments can be expensive and sometimes infeasible. Therefore, a more commonly adopted approach is to develop skillsets and behavioral policies in virtual worlds before deploying in physical environments.

\noindent{\bf (3) Application Tasks and Goals:} correspond to the objectives the agents want to achieve by collaborating with other agents and interacting with environments. 

\noindent{\bf (4) Agents:} include various types of agents. Multiple agents often collaborate for one or more application tasks.  We will discuss this in more detail later.

\noindent{\bf (5) Agent Controller:} coordinates the collaboration and task cognition actions among agents. 
It consists of the following subfunctional modules: (1) {\it Environment and task cognition: } detects the implementation environment and types and requirements of tasks; (2)  {\it Agent adaptation: } coordinates the deployment of agents to fit the needs of specific tasks; and (3) {\it Resource orchestration: } coordinates the allocations of different resources, including communication and computational resources for implementing agents in different environments according to various task requirements.

The development pipeline is described as follows: The agent controller will first identify requested application tasks and the implementation environments. It will then search the library of existing models or agents to see if there exist any models or agents that can be directly deployed to meet the task requirements. The selected models or agents can be further updated or fine-tuned based on the knowledge or data generated by the GF-agents. Finally, the updated agents will be deployed. The performance of the deployed agents will be monitored or inferred based on the dynamics of the task requirement and environment. Once the performance of agents cannot meet the new requirements and/or dynamics of the environment, the above procedures will be repeated. The above pipeline is aligned with the 3GPP's AI/ML development pipeline\cite{Lin20243GPPAIOverview}.


%
%
%
%

\begin{figure}
\centering
\includegraphics[width=3.2 in]{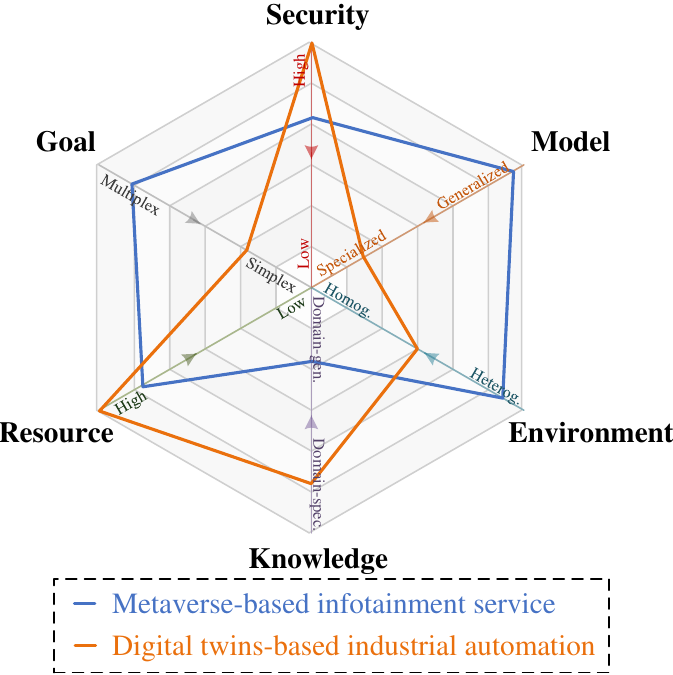}
\caption{{KPI Requirements of two application scenarios, metaverse-based infotainment and digital twins-based industrial automation, on the 6 key performance metrics, measured by model generalization error (Model), non-i.i.d. level of environmental sensing datasets (Environment),  knowledge-associated domain specialty (Knowledge), computational and communication resource cost measured by energy consumptions (Resource), task goal relevant inception scores (Goal), and model bias under adversarial attacks (Security), respectively, all in normalized values.}}
\label{Figure_KPIs}
\end{figure}

\begin{table*}[tbp]
\centering
\caption{Comparison of data-focused communication, existing network AI solutions, and AgentNet.}
\vspace{-0.1in}
\label{Tabel_DataandDLCCompare}
\scriptsize
\begin{tabular}{|l|l|l|l|}
\hline
 & \makecell[c]{\bf Existing Comm. Network} & \makecell[c]{\bf Existing Network AI Solutions} & \makecell[c]{ \bf AgentNet} \\
 \hline
{\bf Basic Idea} & \makecell[l]{Focusing on transporting\\ data packets from one point to another} & \makecell[l]{Pre-selected models trained based on\\ given datasets to extract stationary patterns\\ and insights, or make predictions.} &
\makecell[l]{Autonomous AI networking paradigm focusing on\\ human user-agent-environment interacting, collaborating,\\ and knowledge transfer to achieve various task objectives.}  \\
\hline
\makecell[l]{\bf Key Functional\\ \bf Components} & \makecell[l]{Source and channel\\ encoders and decoders, etc.} & \makecell[l]{Cloud/edge servers, set of model\\ structures,  training datasets} & \makecell[l]{Agents, infrastructure, agent controller,\\ environment, tasks, etc.} \\
\hline
\makecell[l]{\bf Limitations} & \makecell[l]{Task and environment-\\agnostic, low efficiency.} & \makecell[l]{Data transportation and processing\\ separation, low data processing efficiency,\\ high-decision delay, limited flexibility.} & \makecell[l]{Still in the early stage of development.} \\
\hline
\makecell[l]{\bf KPIs} & \makecell[l]{Data rate, bit/symbol-error-rate\\ end-to-end commun. latency, etc.} & \makecell[l]{Communication and computational\\ resource, efficiency, model accuracy, etc.} & \makecell[l]{Model generality, environmental diversity, \\ goal complexity, knowledge/domain generality,\\ security, etc.} \\
\hline
{\bf Use Scenarios} & \makecell[l]{eMBB, URLLC, mMTC} & \makecell[l]{Pattern recognition,  planning, prediction\\ in stationary network environments.} & \makecell[l]{Interactive immersive communication,\\ autonomous network management, etc.}\\ 
\hline
\end{tabular}
\vspace{-0.1in}
\end{table*}
\normalsize

\subsection{Key Performance Metrics}

{In \cite{Shavit2023AgenticAIPractices}, the authors have proposed four performance metrics required for enabling ``agenticness" in a system, including complexity levels of achievable goal and implementable environment of agents as well as the generalization and autonomy level when adapting to unknown and unexpected situations. AgentNet is a special communication networking system consisting of a diverse set of agents that can interact and collaborate for various goals.}  
We extend the concept of ``agenticness" into communication networking systems in which the performance of AgentNet needs to be evaluated based on the following key performance metrics:  

\subsubsection{{(Implementable) Environment-related metrics}} This includes complexity levels of the environments in which agents are deployed to interact and collaborate to achieve their goals. In AgentNet, the complexity of the environment can be measured by a range of metrics, including the diversity, dynamic range, and volatility, as well as the number of domains or modalities of the environment that should be considered to achieve the goal. In Fig. \ref{Figure_KPIs}, we compare environmental complexity based on the diversity levels of local environments of different agents measured by the non-i.i.d levels of locally collected environment-related data for two application scenarios: metaverse-based infotainment (MVI) service and digital twins (DT)-based industrial automation. We can observe that the MVI service generally has a much more complex environment, as the personal demands and requirements of different users may vary across different times and spaces, especially when compared to the DT-based industrial automation in which the environmental complexity must be kept low to ensure high reliability and accuracy\cite{XY2018TactileInternet}.


\subsubsection{Model-related metrics} The selection of models is critical for each agent to deliver the ideal performance when performing specific tasks. 
In AgentNet, each agent can choose a set of foundation models or F-agents for local adaptation to some specific downstream tasks. Some models may be relatively easier to adapt to a wider range of tasks with relatively diverse requirements than others. Here, we borrow terms from meta-learning and refer to these models as ``models with good generalization ability"\cite{Fallah2021Metageneralization}. There may also exist some other models that are relatively easier to adapt to very specialized tasks, which we refer to as the specialized models. 

\subsubsection{Knowledge-related metrics} The service tasks and data samples, as well as models, can be associated with a particular set of knowledge bases, either within a single domain or across multiple domains.
 For instance, domain-specific services are associated with relatively narrow domain knowledge, e.g., services within a particular field, such as industry automation of a very specific manufacturing process. These services are relatively easier to find, simple, and mature models to build agents for data selection, curation, and model training and adaptation, especially compared to the domain-general services, which involve a much wider range of knowledge data samples across different domains.

\subsubsection{Resource-related metrics} The cost of various types of resources, including computational, communication, software, and hardware infrastructure resources consumed in various phases in the agent development pipeline, directly affects the performance of agents. For example, the time duration for training and updating satisfactory models (e.g., with a given model accuracy) for agents is closely related to the 
computational power and storage capacity of the users or edge servers\cite{Zhang2024ZeroEmission}. For collaborative model construction and transfer, the model performance is also related to the communication performance between users. 

\subsubsection{{(Achievable) Goal-related metrics}}
{In AgentNet, different agents are designed to achieve various goals in both physical and virtual world environments. The performance of agents depends not only on high fidelity and accuracy but also on robustness, adaptability, and efficiency. Evaluating the performance of agents based on various aspects of their achievable goals ensures that they can handle complex scenarios and data variability effectively \cite{Deniz2024JSCCProcIEEE}. }




\subsubsection{Security-related metrics}
Recent studies suggest that model-related information, such as parameters and intermediate model training results, can also be exploited to estimate the personal-related information of users. Therefore, novel metrics that measure the security levels of different AgentNet configurations and solutions need to be developed and carefully evaluated. For example, some F-agents can be developed to simulate various attacks on the model and data, as well as the corresponding consequences. These F-agents can then be utilized when developing E-agents to evaluate and choose appropriate security measures when being deployed in different environments.

\section{A Generative Foundation Model-based Implementation Framework and Application Scenarios}
\label{Section_GFMimplementation}

We introduce a generative foundation model (GFM)-based implementation of AgentNet, which relies on GFM-as-agent (GF-agent), a special F-agent pre-trained for generating synthesized data samples and novel contents for a wide range of downstream tasks and goals. These synthesized data samples can be used to simulate the complex interactions and responses of agents when interacting under various seen and unseen scenarios. GF-agents can alleviate the potential bias of human-labeled data samples, and enable rapid adaptation of agents in new scenarios with limited or zero training dataset\cite{XY2022FedSelfSupervised}.  

{In the rest of this section, we focus on a special implementation of GFM-based AgentNet in the cross-layer optimization for a mobile networking system. The AgentNet in this implementation consists of the following types of agents:}

\begin{itemize}
    \item[(1)] {\em Application-layer GF-agent (aGF-agent)}: corresponds to the GF-agent that can directly interact with users and environments through various application interfaces. For example, virtual assistant applications can rely on the large-language model (LLM)-based GF-agent to extract the implicit semantics and goals from the human users' input, e.g., human languages.  In practical implementations, a light-weight E-agent developed based on the aGF-agent can be deployed at the application layer of the UE to interact with human users\cite{XY2023iSAN}.
%
    \item[(2)] {\em Physical-layer GF-agent (pGF-agent)}: corresponds to the GF-agents that can interact and adapt to various physical-layer environments. For example, in mobile networking systems, diffusion model-based pGF agents have shown promising results in estimating, synthesizing, predicting, and optimizing the physical channels connecting UEs and base stations (gNBs) in various environments\cite{Guoxuan2024RFDiffusion}. 
    Our previous work has shown that the environmental semantics, including types of communication environments, e.g., indoor or outdoor, as well as relative locations and directions of transmitters and receivers, can be utilized to construct environment-specific E-agents from a limited number of F-agents\cite{Zhu2024SANSee}. 
    \item[(3)] {\em Network-layer GF-agent (nGF-agent)}: corresponds to GF-agents that can interact with various network-layer environments. For example, our previous work has shown that multi-generator GAN-based nGF-agents can be developed to learn and reproduce complex data distributions, especially those with inherent variability and distinct patterns\cite{XY2022FedSelfSupervised}. In this case, network vulnerabilities and potential risks due to the increase in traffic at some locations can be predicted, and the corresponding adjustments can be recommended to further improve the reliability and robustness of the networking systems. 
\end{itemize}

To verify the performance of our proposed GFM-based implementation, we have developed an AgentNet prototype based on srsRAN, an open source 5G software radio platform, deployed on two computers serving as 1 UE and 1 gNB. The UE and the gNB, each connected to an NI 2944R USRP, communicate with each other via a 5G wireless link. An open 5GS-based 5G core network is also installed. We consider the following two specific application scenarios to discuss the detailed implementation of our proposed GFM-based AgentNet framework. 

\begin{figure}
\centering
\includegraphics[width=3.6 in]{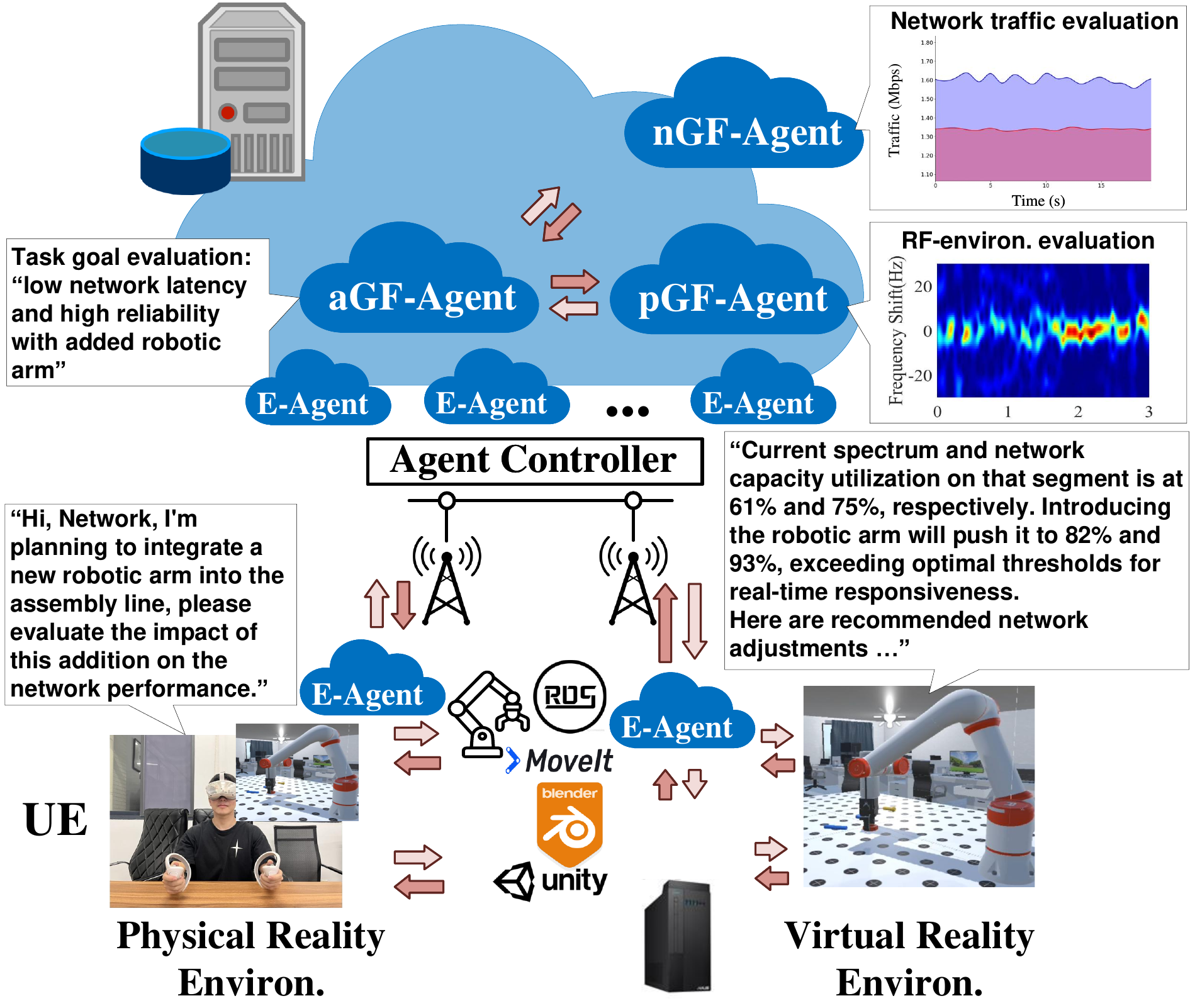}
\caption{{An implementation of AgentNet based on GF-agents for digital twins-based industrial automation: the predicted network traffic with and without the added robotic arm is shown in the right-top subplot, the spectrogram of physical environment caused by the operation of the added robotic arm is shown in the right-middle subplot, and the final evaluation results about physical-layer and network-layer impacts are listed in the text at the right-bottom subplot.}}
\label{Figure_GMFImplementationArchitecture}
\end{figure}
\subsection{Application Scenario 1: Digital Twins-based Industrial Automation}

Digital twin is a key use scenario in 6G that focuses on establishing a virtual world model replicating a specific environment of the physical world. It has been considered one of the powerful tools for accelerating the development, testing, and deployment of novel ideas and solutions within 6G-enabled industrial automation.

%
GF-agent-based AgentNet can be applied to enable dynamic task planning and adaptation to unforeseen events and conditions. We have developed a hardware prototype of the digital twins-based industrial automation based on Franka Emika Panda FR5 robotic arm as illustrated in Fig. \ref{Figure_GMFImplementationArchitecture}. We deploy an LLM-based aGF-agent to detect the key prompts of the human user operator and provide evaluation results and suggestions based on operator's input. The physical environment of our prototype includes a robotic arm deployed in a smart factory, captured by a camera. A virtual reality replica of the physical environment that can simulate the motion planning of the robotic arm is produced based on the Robot Operating System (ROS) and MoveIt's planning pipeline for task-level motion planning. The constructed digital twin scene is streamed to a PICO 4 Pro virtual reality (VR) headset. Human users, i.e., operators, can use the handheld controller connected to PICO 4 Pro to interact with the robotic arm in the digital twin environment. We simulate the case in which a human operator would like to test the task planning and adaptation capability of an assembly line by evaluating the spectrum and network utilization if a new robotic arm is expected to be added to the existing assembly line.
The pGF-agent in our prototype is a diffusion-based channel state information (CSI) prediction model that generates the synthesized CSI signals if a new robotic arm is added. The nGF-agent is based on a multi-generator GAN model that can learn and create synthesized data samples of control data flows generated by the robotic arm sent to the PICO 4 Pro VR headset. Our preliminary simulation results show that our proposed pGF-agent and nGF-agent can successfully learn and predict the spectrum utilization and network traffic of the data flows of the digital twins systems at above 89\% accuracy\cite{XY2022FedSelfSupervised}.

\begin{figure}
\centering
\includegraphics[width=3.6 in]{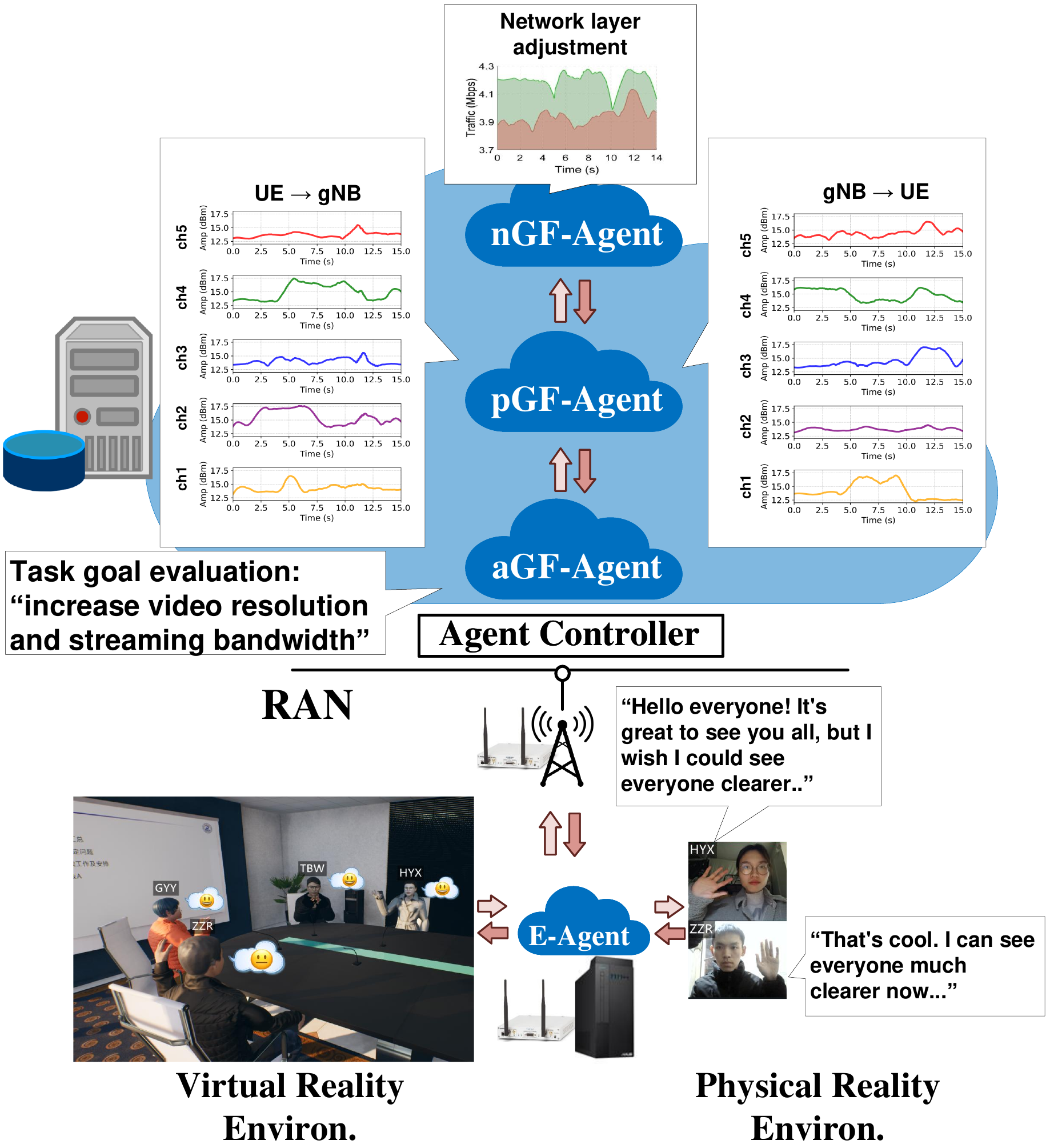}
\caption{{An implementation of GF-agent-based AgentNet for metaverse-based infotainment system: the estimated CSIs in 5 uplink and downlink channels are shown in the middle-left and middle-right subplots, respectively, and the estimated bandwidth required with and without adjusting video resolutions for the network layer is shown in the center-top subplot.}}
\label{Figure_GMFImplementationArchitectureMetaverse}
\end{figure}

\subsection{Application Scenario 2: Metaverse-Based Infotainment System}

We also implement the GF-agent-based AgentNet in a 360-degree video interaction VR prototype, as illustrated in Fig. \ref{Figure_GMFImplementationArchitectureMetaverse}. In this prototype, we adopt the SF3D 2D-to-3D video reconstruction model as the virtual environment creation aGF-agent to create 360-degree multi-view 3D video based on the 2D video of the user captured by a normal computer camera. The converted 3D video is rendered by the Unity engine, which is displayed on a Pico 4 Pro VR headset. An LLM-based aGF-agent is also applied to infer the user's semantics during interaction and then decide the specific steps needed to take to fulfill the user's goal. For example, suppose the LLM-based aGF-agent detects that the user's goal is to increase the resolution of the streamed video, as illustrated in Fig. \ref{Figure_GMFImplementationArchitectureMetaverse}. In this case, the virtual environment creation aGF-agent will first evaluate the data traffic associated with video streaming under different resolutions. A diffusion-based CSI prediction model is implemented in a pGF-agent to predict the CSIs of different frequency bands between the UE and gNB. The pGF-agent will then decide how many more frequency bands should be allocated to support the increased resolution of video streaming. Accordingly, the nGF-agent also needs to evaluate the available bandwidth to make sure the increased traffic can be supported. Our experimental results have shown that, by adopting the GF-agent-based AgentNet, the reliability of video streaming can be guaranteed and the spectrum and network bandwidth utilization can be significantly improved. Notably, by allowing coordination among aGF-agents, pGF-agents, and nGF-agents, the spectrum and network bandwidth utilization can be improved by 46\% and 23\%, respectively, compared to no-coordination case. 

\section{{Open Problems}}
\label{Section_OpenProblem}
\subsection{Agent behavior tracking and malicious action detection}
Since agents are autonomous and can make independent decisions, it becomes difficult for the agent controller to detect suspicious behavior and/or performance degradation in real-time. How to develop a simple and scalable mechanism that can keep track of agent behaviors and automatically trigger an alert when malicious actions are detected is an important research direction for future research.
\subsection{Task-oriented multi-agent coordination mechanism design}
In AgentNet, diverse sets of agents may coordinate, collaborate, and sometimes compete for different task goals in various environments. Developing effective multi-agent coordination mechanisms that support task-oriented interaction, synchronization, and conflict resolution among diverse agents is an interesting problem worth further investigation.
\subsection{Integration with data-oriented networking functions}
AgentNets will not fully replace existing data-oriented communication systems. How to design a hybrid architecture and a compatible interface that allows AgentNet to coexist and jointly work with functional modules and network functions of the existing systems is an important open problem.

\section{Conclusion}
This article has proposed AgentNet, a novel framework for supporting the communication and networking of diverse AI agents. A general architectural framework has been introduced, and a GFM-based implementation has been discussed. Two application scenarios, digital-twin-based industrial automation and metaverse-based infotainment systems, have been discussed, and experimental results have been presented to verify the performance of AgentNet in supporting task-driven collaboration and interaction among AI agents.

\section*{Acknowledgment}
The work of Y. Xiao, G. Shi, and P. Zhang was supported in part by the Mobile Information Network National Science and Technology Key Project under grant 2024ZD1300700. The work of G. Shi was supported in part by the National Natural Science Foundation of China (NSFC) under grant 62293483. The work of P. Zhang was supported in part by the NSFC under grants 62293480 and 62293481.


\bibliographystyle{IEEEtran}
\bibliography{DeepLearningRef}



\begin{IEEEbiographynophoto}{Yong Xiao}(Senior Member, IEEE) is a professor in the School of Electronic Information and Communications at the Huazhong University of Science and Technology (HUST), Wuhan, China. His research interests include semantic communication, green network systems, and Internet-of-Things (IoT).
\end{IEEEbiographynophoto}
\begin{IEEEbiographynophoto}{Guangming Shi} (Fellow, IEEE) 
is the deputy director of Peng Cheng Laboratory, Shenzhen, China. He is also a professor with the School of Artificial Intelligence, Xidian university. 
His research interest includes artificial intelligence, semantic communication, brain-inspired computing, and signal processing.  
\end{IEEEbiographynophoto}
\begin{IEEEbiographynophoto}{Ping Zhang} (Fellow, IEEE) 
is currently a Professor with the School of Information and Communication Engineering, BUPT, and also the Director of the State Key Laboratory of Networking and Switching Technology. 
He is an Academician with the Chinese Academy of Engineering (CAE). His research is in the board area of wireless communications. 
\end{IEEEbiographynophoto}

\end{document}